\documentclass[aps,prl,showpacs,twocolumn,groupedaddress]{revtex4}  
\usepackage[latin1]{inputenc} 
\usepackage{graphicx,epsfig,epstopdf}  
 
\begin{document} 
 
\title{Multiple Magnon Modes and Consequences for the Bose--Einstein Condensed Phase in  
BaCuSi$_{2}$O$_{6}$} 

\author{  
Ch. R\"uegg,$^{1,*}$ 
D. F. McMorrow,$^{1,2}$ 
B. Normand,$^{3}$ 
H. M. R$\o$nnow,$^{4}$ 
S. E. Sebastian,$^{5}$ 
I. R. Fisher$^{5}$ 
C. D. Batista,$^{6}$ 
S. N.~Gvasaliya,$^{4}$ 
Ch. Niedermayer,$^{4}$ 
and J. Stahn$^{4}$} 
 
\email{c.ruegg@ucl.ac.uk} 
 
\affiliation{  
$^1$ London Centre for Nanotechnology; Department of Physics and Astronomy;  
University College London; London WC1E 6BT; UK \\  
$^2$ ISIS Facility; Rutherford Appleton Laboratory; Chilton; Didcot  
OX11 0QX; UK \\ 
$^3$ Institute of Theoretical Physics; Ecole Polytechnique F\'ed\'erale  
de Lausanne; 1015 Lausanne;  
Switzerland \\  
$^4$ Laboratory for Neutron Scattering; ETH Zurich and Paul Scherrer  
Institute; 5232 Villigen PSI; Switzerland\\  
$^5$ Geballe Laboratory for Advanced Materials and Department of Applied  
Physics; Stanford University; Stanford CA 94305; USA \\ 
$^6$ Condensed Matter and Statistical Physics; Los Alamos National  
Laboratory; Los Alamos; USA}  
 
\date{\today} 
 
\begin{abstract} 
The compound BaCuSi$_{2}$O$_{6}$ is a quantum magnet with antiferromagnetic  
dimers of $S = 1/2$ moments on a quasi--2D square lattice. We have  
investigated its spin dynamics by inelastic neutron scattering experiments  
on single crystals with an energy resolution considerably higher than in  
an earlier study. We observe multiple magnon modes, indicating clearly the  
presence of magnetically inequivalent dimer sites. This more complex spin  
Hamiltonian leads to a distinct form of magnon Bose--Einstein  
condensate (BEC) phase with a spatially modulated condensate amplitude.  
\end{abstract} 
 
 
\maketitle 
 
The investigation of field--induced quantum phase transitions (QPTs) in  
magnetic insulators continues to enrich our understanding of the possible  
quantum ground states of matter \cite{Affleck91, Giamarchi99, Sachdev99,  
Nikuni00, Rice02}. Structurally dimerized quantum spin systems play a  
preeminent role in these studies. One recent example is the phase diagram  
of the compound BaCuSi$_2$O$_6$ \cite{Sasago97, Jaime04, Sebastian05,  
Sebastian06}, in which antiferromagnetic dimers formed by the $S = 1/2$  
magnetic moments from pairs of Cu$^{2+}$ ions are arranged on a  
quasi--two--dimensional (2D) square lattice. This material is one of  
the best candidates for the investigation of field--induced BEC of  
magnetic quasiparticles, and a 3D--2D dimensional crossover at the QPT  
has been reported \cite{Sebastian06}. 
 
The ground state of weakly interacting antiferromagnetic dimers is a  
spin singlet ($|S,S_z \rangle = |0,0 \rangle$) separated by an energy gap  
$\Delta$ from excited triplet states ($|1,0 \rangle$ and $|1, \pm1 \rangle$).  
A QPT occurs at the field $H_c = \Delta / g \mu_B$, where the $S_z = +1$  
component condenses. At $H > H_c$ the average triplet density becomes  
finite (magnon BEC) and the ground state changes from a nonmagnetic singlet  
to an ordered magnetic phase. The triplet quasiparticles may also be 
considered as hard--core bosons with a kinetic energy and an effective 
repulsion. Depending on the balance between these terms, a characteristic 
spatial modulation of the triplet density may arise, whereas in a uniform BEC 
this is identical for all sites. 
 
\begin{figure}[t] 
\includegraphics[width=0.42\textwidth]{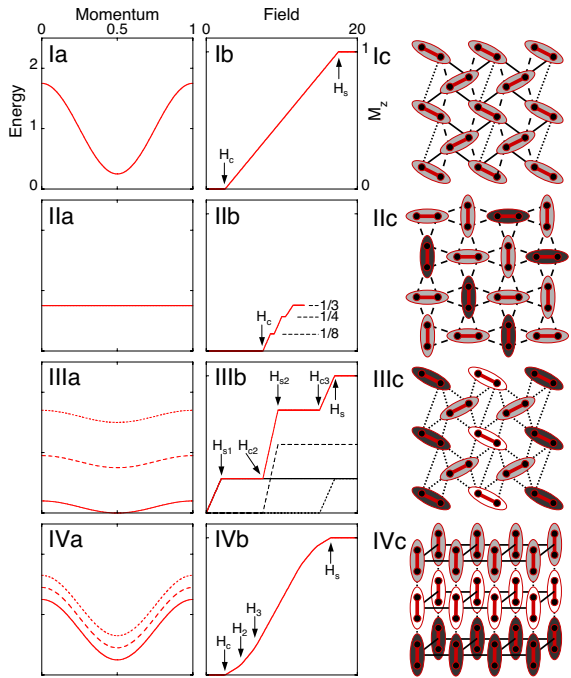}  
\caption{\small Physical properties at a field--induced QPT in  
dimer--based $S = 1/2$ spin systems. (a) triplet dispersion at  
zero field, (b) uniform magnetization $M_{z}$, (c) dimer lattice with  
triplet condensate amplitude for $H > H_c$ represented by increasing  
grey scale. (I) 3D systems with uniform magnon BEC, e.g.~TlCuCl$_3$.  
(II) Shastry--Sutherland geometry with magnetization plateaus and strong  
contrast in condensate amplitude, e.g.~SrCu$_2$(BO$_3$)$_2$ (field in  
panel (c) yielding $M_z = 1/3$). (III) Same features for 3D system with  
strong translational symmetry--breaking and weak interdimer interactions,  
e.g.~NH$_4$CuCl$_3$ at $H_{c2} < H < H_{s2}$. (IV) Quasi--2D system on  
square lattice with weakly broken translational symmetry showing no  
magnetization plateaus, e.g.~schematic model for BaCuSi$_2$O$_6$ at  
$H_{c} < H < H_{3}$.} 
\label{fig4}  
\end{figure}

The classes of field--induced QPT known to date in dimer spin systems are  
summarized in Fig.~\ref{fig4}. For magnetic interactions with weak or no  
frustration, the kinetic energy is dominant and the ordered, or  
BE--condensed, ground state is uniform at $H > H_c$, as in TlCuCl$_3$  
\cite{Rice02, Nikuni00, Rueegg03, Matsumoto02}. In SrCu$_2$(BO$_3$)$_2$  
\cite{Rice02, Kageyama99, Kodama02} triplet hopping is suppressed by  
geometrical frustration and the repulsion causes the condensed triplets  
to form a superlattice with spontaneous breaking of translational symmetry  
and the appearance of magnetization plateaus \cite{Oshikawa97}. All dimer  
sites have finite, if weak, triplet density at $H > H_c$ \cite{Kodama02}.  
Magnetization plateaus occur also for strong and explicit translational  
symmetry breaking, which leads to inequivalent dimer sites and multiple  
magnon modes, as in NH$_4$CuCl$_3$ \cite{Matsumoto03, Rueegg04} where the  
separation of the singlet--triplet gaps exceeds the magnon band widths.  
 
Prior to this study the magnetic Hamiltonian used to describe  
BaCuSi$_2$O$_6$ was based on an inelastic neutron scattering (INS)  
investigation with coarse energy resolution \cite{Sasago97} and on  
fits to thermodynamic data \cite{Jaime04, Sebastian05}. Motivated by  
the exotic low--temperature behavior \cite{Sebastian06}, the lack of  
high--resolution data and reports of a structural phase transition  
around 100 K \cite{Sparta04, Samulon06}, we have investigated the
magnetic excitation spectrum by high--resolution INS. We find explicitly 
broken translational symmetry in the form of structurally inequivalent 
dimer sites, which produces a BEC with spatial modulation of the  
triplet density, but without magnetization plateaus. As a consequence 
our conclusions extend beyond the determination of the spin Hamiltonian 
to demonstrate that BaCuSi$_2$O$_6$ represents a fourth category  
[Fig.~\ref{fig4}(IV)] of spatially modulated BEC system, and we  
propose a closer examination of the magnetic properties around the  
field--induced quantum critical point (QCP). 
 
\begin{figure}[t!]  
\includegraphics[width=0.42\textwidth]{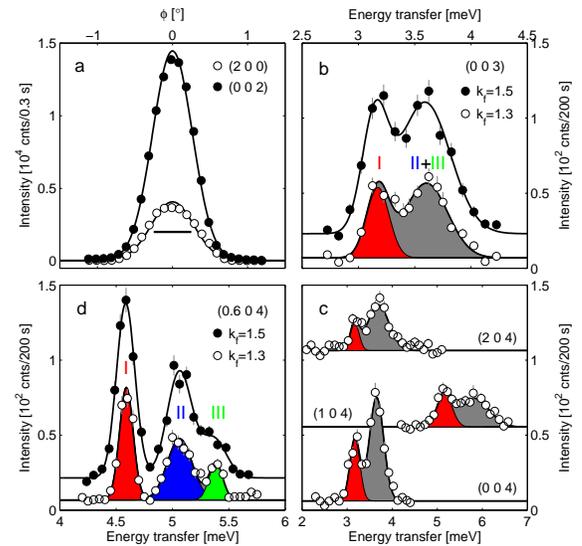}  
\caption{\small (a) Rocking curves across two Bragg peaks of coaligned  
BaCuSi$_2$O$_6$ sample. Horizontal line indicates instrumental resolution  
(full width) of neutron diffractometer MORPHEUS. (b--d) INS spectra of  
BaCuSi$_2$O$_6$ at $T = 1.8$ K and $H = 0$ T, experimental  
conditions as indicated. Momentum vectors in reciprocal lattice units  
(r.l.u.) of tetragonal space group $I4_{1}/acd$ with $a = b = 10.01$ \AA~and  
$c = 22.47$ \AA~\cite{Sparta04}, i.e.~($q_{h}$~0~4) is diagonal direction on  
square lattice.}  
\label{fig1}  
\end{figure}

Five single crystals of BaCuSi$_2$O$_6$ were coaligned to provide a total  
mass of 1.25 g. The total mosaic spread of the assembly was excellent and  
matched the instrumental resolution [Fig.~\ref{fig1}(a)]. A  
tetragonal--to--orthorhombic distortion occurring below  
$T\approx100$ K was detected very recently by high--resolution x--ray  
diffraction \cite{Samulon06}. The small difference between $a$-- and $b$--axes 
is related to a structural modulation with wave vector (0~0.129~0); crystals  
are twinned below this transition \cite{Samulon06}, which should result in  
spatial averaging of in--plane properties.  
 
INS spectra at $T = 1.8$ K and zero magnetic field were collected on the  
cold--neutron triple--axis spectrometer TASP \cite{Semadeni01}, operated  
with fixed final momentum $k_f = 1.3$ \AA$^{-1}$ ($1.5$ \AA$^{-1}$),  
focusing Pyrolytic Graphite monochromator and analyzer, cold Be filter, and  
open horizontal collimation to gain intensity. The Gaussian energy resolution  
of 0.14(0.23) meV (FWHM) is considerably improved compared to that in  
Ref.~\cite{Sasago97}. Figure \ref{fig1}(b) shows the spectrum at the AF  
zone center \boldmath$Q$\unboldmath$=$($q_{h}$~$q_{k}$~$q_{l}$)$=$(0~0~3) and  
can be compared directly to Fig.~2(b) of Ref.~\cite{Sasago97}. Surprisingly,  
the singlet--triplet gap has a clear multi--peak structure over an energy  
range of 1 meV. This splitting cannot be attributed to the multi--crystal  
sample [Fig.~\ref{fig1}(a)], being observed consistently in different  
instrumental configurations including flat analyzer, and confirmed on  
different instruments (below). We conclude that the triplet excitation  
spectrum of BaCuSi$_2$O$_6$ shows intrinsically more features than  
originally reported. 
 
The INS spectrum in Fig.~\ref{fig1}(b) is described adequately by two 
Gaussian peaks. However, while the excitation at lower energy transfer, 
denoted henceforth as mode I, has an energy width compatible with the 
instrumental resolution, the peak at higher energies is considerably 
broader. The evolution of the inelastic signal in the square--lattice 
plane is presented in Fig.~\ref{fig1}(c), and the multi--peak structure 
is observed at every point. At (0.6~0~4) the slope of the resolution 
ellipsoid matches that of the dispersion relation, giving optimal 
resolution conditions [Fig.~\ref{fig1}(d)]: this spectrum reveals that 
the excitation at higher energy transfer consists of at least two 
transitions, which are denoted as modes II and III. However, peak II 
remains broader in energy than expected from the instrumental resolution 
(cf.~mode I), and therefore may consist of more than one mode. 
 
\begin{figure}[t]  
\includegraphics[width=0.42\textwidth]{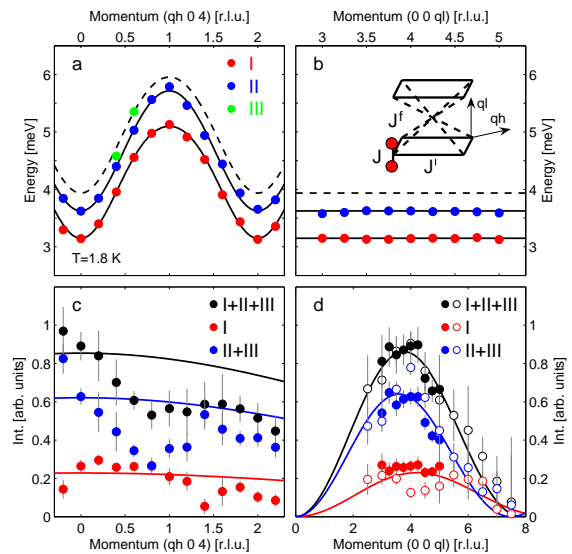}  
\caption{\small (a--b) Triplet dispersion in BaCuSi$_2$O$_6$ at zero field  
and $T = 1.8$ K: (a) in--plane along \boldmath$Q$\unboldmath$=$($q_{h}$~0~4),  
(b) perpendicular to plane with \boldmath$Q$\unboldmath$=$(0~0~$q_{l}$) (see  
inset). (c--d) Momentum--dependence of (scaled) INS intensity, with solid  
lines as described in text: total intensity modes I--III (black), mode I  
(red), and modes II+III (blue) measured with $k_f = 1.3$ \AA$^{-1}$ (solid  
circles) and $1.5$ \AA$^{-1}$ (open circles).}  
\label{fig2}  
\end{figure}

The energy dispersion of modes I--III is extracted from the data by  
least--squares fits of the resolved peaks, and is summarized in  
Figs.~\ref{fig2}(a) and (b). A cosine dispersion is observed in  
the square--lattice plane: modes I--III disperse in parallel and  
are well described by the lowest--order perturbative expression  
\begin{eqnarray} 
E_{\alpha}({\mbox{\boldmath$Q$\unboldmath}}) = J_{\alpha} & - & J'_{\alpha} 
(\cos(\pi q_h + \pi q_k) + \cos(\pi q_h - \pi q_k)) \label{efe} \nonumber\\ 
& + & J''_{\alpha}(\cos(2\pi q_h) + \cos(2\pi q_k))\\ 
& + & 2 J_\alpha^{f} \cos ( \textstyle{\frac{\pi}{2}} q_l ) 
|\cos(\pi q_h) - \cos(\pi q_k)| \nonumber,  
\end{eqnarray} 
where $\alpha = 1,2,3$ denotes modes I--III. Parameters $J_{\alpha}$  
are intradimer interactions, with $J'_{\alpha}$ and $J''_{\alpha}$  
respectively nearest-- and next--nearest--neighbor in--plane interdimer  
interactions and $J_\alpha^{f}$ the effective interlayer interactions.  
The measured energies have no detectable $q_{l}$--dependence (bandwidth  
below 0.05 meV), confirming previous claims that BaCuSi$_2$O$_6$ is quasi--2D. 
This is a consequence of the frustrated geometry of the interlayer coupling  
[inset Fig.~\ref{fig2}(b)], and the precise determination of $J_\alpha^f$  
is deferred to a future experiment. Here we take $J_\alpha^{f} = 0$,  
consistent with the fact that the other parameters in Eq.~(\ref{efe})  
correspond to decoupled planes and no more general model is required.  
A fit to the complete data set gives dominant intradimer interactions  
$J_{1} = 4.27(1)$, $J_{2} = 4.72(1)$, and $J_{3} = 5.04(4)$, with $J'_{1}  
= 0.49(1)$, $J'_{2} = 0.52(1)$, $J''_{1} = - 0.07(1)$, and $J''_{2} =  
- 0.03(1)$, all in meV; we take $J'_{3} = {\textstyle \frac{1}{2}} 
(J'_{1} + J'_{2})$ and $J''_{3} = {\textstyle \frac{1}{2}} (J''_{1} +  
J''_{2})$ to approximate the dispersion of mode III. Thus we find in  
contradiction to Ref.~\cite{Sasago97} that the only significant interdimer  
interactions are those forming the square lattice, $J'_{\alpha} \approx  
0.1 J_{\alpha}$. The individual gaps at the AF zone center are $\Delta_{1}  
= 3.15(5)$, $\Delta_{2} = 3.62(5)$, and $\Delta_{3} = 3.94(8)$ meV.  
 
The total INS intensity of modes I--III is modulated along (0~0~$q_{l}$)  
by the dimer structure factor $|f($\boldmath$Q$\unboldmath$)|^2 \sin  
(d q_{l})^{2}$ \cite{Sasago97} [Fig.~\ref{fig2}(d)], with $2d =  
2.85(6)$ \AA~the average projected intradimer separation and  
$f($\boldmath$Q$\unboldmath$)$ the magnetic form factor of Cu$^{2+}$.  
However, fitting intensities I and II+III separately gives $2d_I = 2.52(18)$  
\AA~ and $2d_{II+III} = 2.99(10)$ \AA~[Fig.~\ref{fig2}(d)], a difference  
indicative of inequivalent dimer layers, as already suggested by the  
multi--mode dispersion. However, a model with uniform layers of dimer oriented 
parallel to the $c$--axis and modulated only in this direction would predict  
a weakly decreasing INS intensity with increasing $|$($q_{h}$~0~4)$|$. The  
$q_{h}$--dependence of the measured intensity [Fig.~\ref{fig2}(c)] shows  
further structure which is clearly inconsistent with equivalent dimer  
sites in each plane [and thus also with the simple representations in  
Fig.~\ref{fig4}(IVc) and Eq.~(\ref{efe})]. We are forced to conclude that  
the magnetic interactions are also modulated within the square planes,  
consistent with what is known of the low--temperature structure  
\cite{Samulon06}, as well as between adjacent planes. 
 
\begin{figure}[t] 
\includegraphics[width=0.42\textwidth]{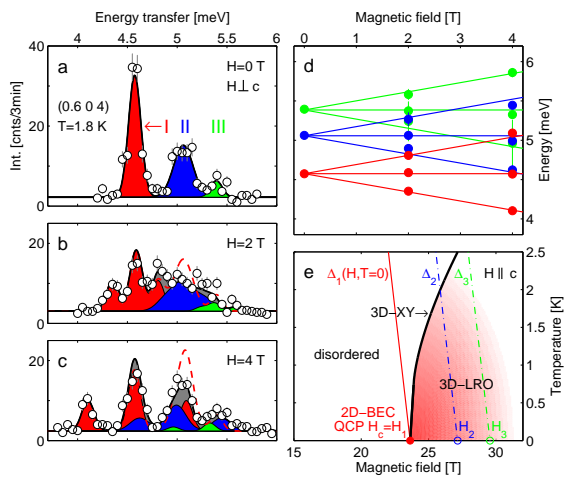}  
\caption{\small (a--c) INS spectra at three magnetic--field values in  
BaCuSi$_2$O$_6$ for \boldmath$Q$\unboldmath$=$(0.6~0~4), $T = 1.8$ K,  
and $k_{f} = 1.3$ \AA$^{-1}$. 
(d) Zeeman splitting of triplet modes following procedure described in  
text. (e) Schematic phase diagram around QPT; intensity of red shading  
indicates qualitative degree of $c$--axis modulation of BEC order  
parameter.}  
\label{fig3}  
\end{figure}

The dependence of the excitation spectrum on magnetic fields up  
to $H = 4$ T is shown for the resolution--focusing point in Fig.~\ref{fig3}.  
These data were collected on the spectrometer RITA--II \cite{Lefmann00} with  
experimental conditions similar to TASP. 
The $H = 0$ T data in Fig.~\ref{fig3}(a) confirm our conclusions concerning  
the multi--peak excitation structure [cf.~Fig.~\ref{fig1}(d)]. At finite  
magnetic fields a redistribution of spectral weight is observed. While mode  
I is clearly split, the transitions at higher energies are barely resolved  
due to peak overlap [Figs.~\ref{fig3}(b,c)]. One may proceed by fixing the  
peak widths at the zero--field values and redistributing the spectral weight  
according to the relative contributions expected for Zeeman--split triplet  
modes, namely 1/4, 1/2, and 1/4 respectively for $S_z = +1, 0$, and $-1$  
components; the only fitting parameters are the centers of the Gaussian  
peaks. This approach describes the observed finite--field spectra quite  
satisfactorily, whereas a model with field--independent modes II and III  
cannot [dashed lines in Figs.~\ref{fig3}(b,c)]. Fits of individual energies,  
${\tilde E}_{\alpha}$, and a common $g$--factor are  
summarized in Fig.~\ref{fig3}(d): ${\tilde E}_{1} = 4.58(1)$ meV,  
${\tilde E}_{2} = 5.06(1)$ meV, and ${\tilde E}_{3} = 5.39(1)$ meV  
[cf.~Fig.~\ref{fig1}(d)], while $g = 2.01(4)$, in good agreement with the  
value reported in Ref.~\cite{Sebastian05}. That each of the modes I--III  
displays individual Zeeman splitting into components $E_\alpha^{\pm,0} (H)$  
demonstrates again the presence of inequivalent dimer sites. Different  
candidate mechanisms for a zero--field energy splitting, such as exchange  
anisotropy or Dzyaloshinskii--Moriya interactions \cite{Glazkov04, Rueegg03},  
are probably present in BaCuSi$_2$O$_6$ at some small energy scale \cite{Sebastian06ESR}. However,  
these would produce only three modes at finite fields and are therefore  
excluded as the origin of our observations.  
 
The INS results presented here identify BaCuSi$_2$O$_6$ in a class of  
low--dimensional quantum magnets which has not yet been considered: despite  
explicitly broken translational symmetry of the spin Hamiltonian there are  
no plateaus in the uniform magnetization (Fig.~\ref{fig4}) \cite{Jaime04}.  
Both the presence of four dimer layers along the $c$--axis of the large  
unit cell and the structural distortion at $T \approx 100$ K introduce the  
potential for different intra-- and interdimer interactions. Our results  
indicate that both interlayer and weak in--plane modulations are present,  
and that these are due most likely to differential dimer tilts away from  
the $c$--axis. A minimal magnetic model could be expected to generate at  
least 4 excitations, consistent with the broad, unresolved appearance of  
mode II.  
 
The inequivalent dimer sites reflected in the INS spectra result in gaps  
$\Delta_{\alpha}$ which differ by less than the band widths of the  
individual triplets, in contrast to NH$_4$CuCl$_3$ \cite{Rueegg04} 
[Figs.~\ref{fig4}(IIIa,IVa)], removing the possibility of magnetization  
plateaus [Figs.~\ref{fig4}(IIIb,IVb)]. Each of the measured modes shows  
Zeeman splitting in an applied field, which would drive the spin system 
towards a QCP occurring when the lowest gap $\Delta_{1}$ vanishes
at $H_{1} = H_c$. Above $H_{c}$, the density of triplet quasiparticles  
would be largest for sites of type $\alpha = 1$, but because of their direct  
real--space coupling, dimers of types $\alpha = 2$ and 3 would also acquire  
finite, if weak, triplet densities \cite{Kodama02, Matsumoto03}. The BEC  
state must therefore be characterized by a spatially modulated amplitude 
of the order parameter. As the field is increased through the values  
$H_\alpha$ corresponding to the gaps $\Delta_\alpha$ ($\alpha = 2$ and 3),  
raising the uniform magnetization and the BEC order parameter, the effect  
of interdimer interactions in a system such as BaCuSi$_2$O$_6$ is to cause  
a mixing of the low--lying triplet modes $E_\alpha^{+} (H)$ with the ground  
state. This mixing precludes further phase transitions and causes the  
magnetization to display a crossover rather than a kink at these fields  
[Fig.~\ref{fig4}(IVb)], in accord with the measurement of Ref.~\cite{Jaime04}. 
The spatial modulation of the BEC order parameter may in principle be  
detected by nuclear magnetic resonance (NMR) studies of the many  
inequivalent Cu sites at $H > H_{c}$ \cite{Horvatic05}. 
 
In BaCuSi$_2$O$_6$ our results indicate that the dimer modulation occurs  
predominantly between planes [Figs.~\ref{fig2}(c,d)], and thus that the  
indices $\alpha$ correspond to separate bilayers. In this case the triplet  
density at $H_{c} < H < H_{2}$ is higher on dimer layers $\alpha = 1$, and 
is lower on the intervening bilayers [cf.~Fig.~1(IVc)]; indeed the structural 
modulation must lead in this way to an enhancement of the anisotropic  
nature of the BEC order parameter, as depicted in Fig.~\ref{fig3}(e). 
The structural modulation also raises the distinct possibility that  
geometrical frustration of interlayer triplet hopping [Eq.~(\ref{efe})]  
is less than perfect at low temperatures. However, the thermodynamic  
measurements of a 2D QCP in BaCuSi$_2$O$_6$ as a consequence of this  
frustration \cite{Sebastian06} demonstrate that the energy scale of  
any unfrustrated hopping component is below 30 mK. 
 
In summary, we have investigated the magnetic excitation spectrum of  
BaCuSi$_2$O$_6$ in zero field and at $H < H_{c}$ by inelastic neutron  
scattering on single crystals. The very much higher energy resolution  
than in earlier studies allows the determination of a spin Hamiltonian  
whose exchange interactions indicate a complex picture of the ground  
state and of the field--induced QPT observed in this material. 
Inequivalent dimer sites give a multi--mode excitation spectrum, and 
as a further consequence the BEC phase at $H > H_c$ is anticipated to be spatially 
modulated. Although INS studies are currently limited to 17 T $< H_{c}$,  
high--field magnetization and NMR experiments could explore in greater  
detail the modulated BEC phase whose characteristic features are predicted  
here. 
 
We thank N. Harrison, M. Jaime and R. Stern for valuable discussions. This  
work is based on experiments performed at the Swiss spallation neutron source  
SINQ, Paul Scherrer Institute, Villigen, Switzerland. The project was  
supported by the Swiss National Science Foundation and by the US National  
Science Foundation, Division of Materials Research, under grant DMR-0134613.

\end{document}